\documentclass[runningheads]{llncs}
\usepackage{color}
\usepackage{graphicx}
\begin{document}
\title{AutoPET Challenge 2022: 
Automatic Segmentation of Whole-body Tumor Lesion Based on Deep Learning and FDG PET/CT}
\titlerunning{AutoPET Challenge 2022}
%
\author{Shaonan Zhong\orcidID{0000-0003-1648-1489} \and 
Junyang Mo\orcidID{0000-0002-2981-9349} \and 
Zhantao Liu\orcidID{0000-0002-3934-2660}}
\authorrunning{Zhong et al.}
%
\institute{School of Biomedical Engineering, 
Shenzhen University, Shenzhen Guangdong 518055, China}
\maketitle            
\begin{abstract}
Automatic segmentation of tumor lesions is a critical initial processing step for quantitative PET/CT analysis. However, numerous tumor lesion with different shapes, sizes, and uptake intensity may be distributed in different anatomical contexts throughout the body, and there is also significant uptake in healthy organs. Therefore, building a systemic PET/CT tumor lesion segmentation model is a challenging task. In this paper, we propose a novel training strategy to build deep learning models capable of systemic tumor segmentation. Our method is validated on the training set of the AutoPET 2022 Challenge. We achieved 0.7574 Dice score, 0.0299 false positive volume and 0.2538 false negative volume on preliminary test set.The code of our work is available on the following link: https://github.com/ZZZsn/MICCAI2022-autopet. 

\end{abstract}
\section{Introduction}
Positron Emission Tomography/Computed Tomography (PET/CT) is an integral part of the diagnostic workup for various malignant solid tumor entities. Due to its wide applicability, Fluorodeoxyglucose (FDG) is the most widely used PET tracer in an oncological setting reflecting glucose consumption of tissues, e.g. typically increased glucose consumption of tumor lesions.

As part of the clinical routine analysis, PET/CT is mostly analyzed in a qualitative way by experienced medical imaging experts. Additional quantitative evaluation of PET information would potentially allow for more precise and individualized diagnostic decisions.

A crucial initial processing step for quantitative PET/CT analysis is segmentation of tumor lesions enabling accurate feature extraction, tumor characterization, oncologic staging and image-based therapy response assessment. Manual lesion segmentation is however associated with enormous effort and cost and is thus infeasible in clinical routine. Automation of this task is thus necessary for widespread clinical implementation of comprehensive PET image analysis.

Recent progress in automated PET/CT lesion segmentation using deep learning methods has demonstrated the principle feasibility of this task. However, despite these recent advances tumor lesion detection and segmentation in whole-body PET/CT is still a challenging task. The specific difficulty of lesion segmentation in FDG-PET lies in the fact that not only tumor lesions but also healthy organs (e.g. the brain) can have significant FDG uptake; avoiding false positive segmentations can thus be difficult.

In this study, we first propose a 2D U-Net based whole-body training strategy to address the challenge of whole-body tumor lesion segmentation.

\section{Method}

\subsection{Whole-body training strategy} 

\begin{figure}
\includegraphics[width=\textwidth]{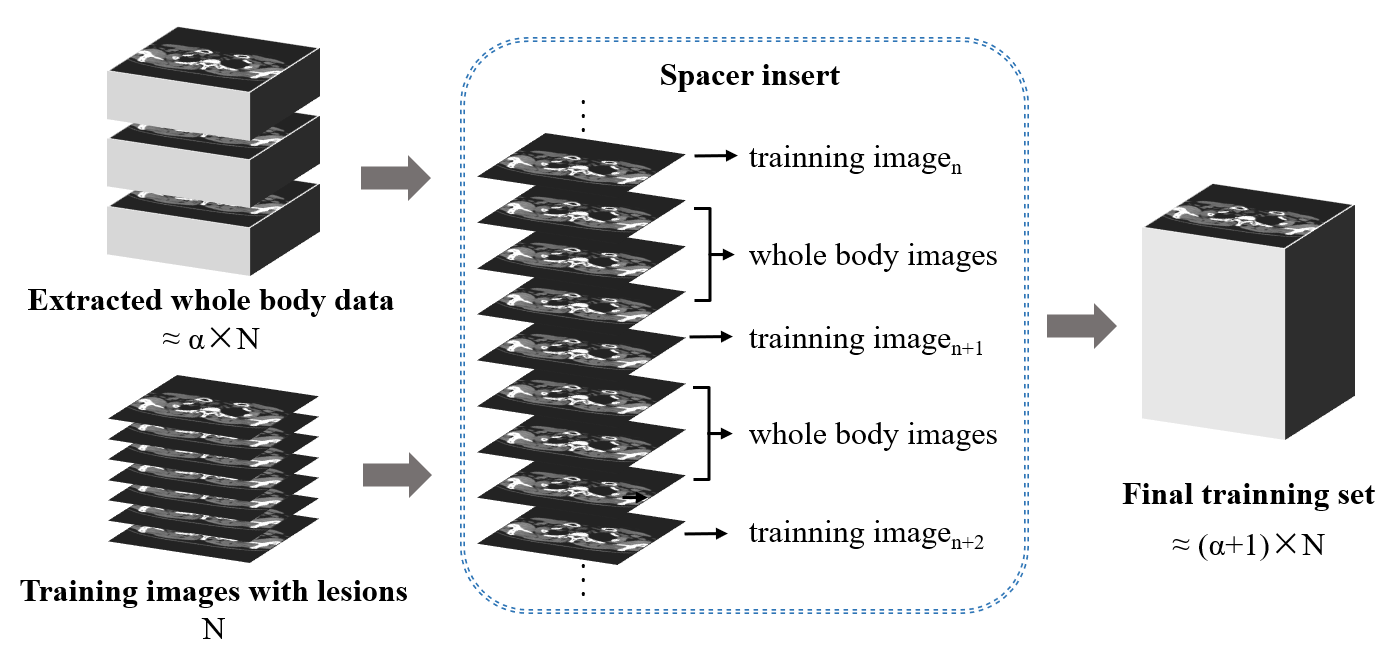}
\caption{Schematic diagram of the whole-body data-extraction strategy. N is the number of original training sets.} \label{fig1}
\end{figure}

Traditional deep learning strategy often only use data containing targets for training, but models trained only using images containing lesions cannot learn the difference between lesions and normal physiological metabolism in the whole body. While, directly using whole body images for training will cause a serious imbalance in the ratio of the lesion layer to the normal layer in the training data, making it difficult for the model to learn to recognize the target lesion. On the basis of training and validation using only images containing lesions, we design a new training strategy that can balance the use of other images without lesions to supplement the information of whole body.

As shown in \textcolor{red}{Fig. 1}, we define a whole-body proportion coefficient $\alpha$, which determines the ratio of the number of images containing lesions to the number of whole-body images. Calculate the number of patients corresponding to the number of whole-body images that need to be added based on the whole-body proportion coefficient and the whole-body slices of the patient images. Then randomly extract the whole body data of the corresponding number of patients in the training set and insert them at proportional intervals into the training data that originally only contained images of lesions.

In addition, in order to reasonably utilize the whole-body data of each patient in the training set, we design a method to use the whole-body data cyclically: define a cyclic epoch $\beta$, after the model training for $\beta$ epochs in the above final training set, extract the whole-body data of the next batch to the original training set to form a new final training set, and then train for $\beta$ epochs again. Through this rotation, we input the whole-body data of each patient in the training set into the model for learning.

In order to prevent the network getting totally different information between two batches of whole-body data, we use a randomly overlap strategy: randomly extracting 25$\%$ data of the first batch to the second batch, and so on, and 25$\%$ data of the final batch to the first batch.

\subsection{Model structure}
Our model is modified based on U-Net \cite{ref_article1}, changing the input of the model to two-channel (PET/CT) and replacing the U-Net encoder with RegNet \cite{ref_article2}. We choose the model structure of RegNetY-16GF they provided for replacement, whose encoder is composed of 4 stages, and each stage consists of a series of block stacks. Each block is a residual structure with grouped convolutions, and each convolution is followed by a Squeeze-and-Excitation module \cite{ref_article3}, so that the model has a self-attention mechanism.

\subsection{Training and hyperparameter optimization}
We used all data of patients whose labels are not negative, and randomly split the data into 5 parts. One part is used as the validation set, and the rest is used as the training set. The PET images were normalized to standard uptake values (SUV). Subsequently, we cropped the cross-section of the PET/CT images to 256×256 to get rid of the most of blank area and make the size suitable for our network.

The network model was built with Pytorch and trained on an Nvidia TITAN XP with 12 GB of memory. Setting the parameters of the whole body training strategy, $\alpha$ is 1 and $\beta$ is 5. In the training phase, we train based on the proposed training strategy, using a batch size of 24 for 80 epochs. Using the AdamW optimizer, the $\beta$1 and $\beta$2 were set to 0.5, and 0.999, respectively. The learning rate adjustment strategy was cosine annealing, the initial learning rate was 10\textsuperscript{-5}, and the final learning rate was reduced to 10\textsuperscript{-8} within 
100 epochs. Saving the model with the best performance on the validation set. In terms of loss function, we choose the combination of Dice loss, Lovasz loss and Binary Cross Entropy (BCE) loss. The weight of each loss function is learned automatically by a network with learning rate set to 10\textsuperscript{-4}.

\section{Result}

\begin{table}
\caption{Segmentation Result of our method.}\label{tab1}
\centering
\begin{tabular}{cccc} 
\hline
                                  & \textbf{~ ~DSC~ ~} & \textbf{~ ~False Negative~ ~} & \textbf{~ ~False Negative~ ~}  \\ 
\hline
\textbf{Validation set}           & 0.6863  & 5.2258 & 7.4603    \\
\textbf{~ Preliminary~Test set~~} & 0.7574               & 0.0299                           & 0.2538                            \\
\hline
\end{tabular}
\end{table}

\section{Conclusion}
We develop a deep learning approach to address the challenge of systemic cancer lesion segmentation in FDG PET/CT scans.


\begin{thebibliography}{3}
\bibitem{ref_article1} Ronneberger O, Fischer P, and Brox T. U-net: Convolutional networks for biomedical image segmentation[C]//International Conference on Medical image computing and computer-assisted intervention. Springer, Cham, 2015: 234-241.
\bibitem{ref_article2} Radosavovic I., Kosaraju R. P., Girshick R., He K., and Dollár P. Designing network design spaces[C]//Proceedings of the IEEE/CVF Conference on Computer Vision and Pattern Recognition. 2020: 10428-10436.
\bibitem{ref_article3} Hu J, Shen L, and Sun G. Squeeze-and-excitation networks[C]//Proceedings of the IEEE conference on computer vision and pattern recognition. 2018: 7132-7141.

\end{thebibliography}
\end{document}